\documentclass[12pt]{article}
\usepackage[all]{xy}
\usepackage{amsmath}[1996/11/01]
\usepackage{amssymb,amsthm,amsfonts,latexsym,epsfig,graphics}
 \def\beql#1#2\eeql{\begin{equation}\label{#1}#2\end{equation}}

\advance \textheight by -1 \topmargin
\advance \textheight by -1 \headheight
\advance \textheight by -1 \headsep
\advance \textheight by -2 \footskip
\vsize = \textheight
\hsize = \textwidth
\DeclareMathOperator{\syn}{{\sc syn}}

\DeclareMathOperator{\Gr}{\mathfrak Gr}
\DeclareMathOperator{\Fl}{\mathfrak Fl}
\DeclareMathOperator{\frk}{frk}
\DeclareMathOperator{\rk}{rk}

\DeclareMathOperator{\pr}{pr}
\DeclareMathOperator{\GL}{GL}

\DeclareMathOperator{\diag}{diag}

\DeclareMathOperator{\id}{id}

\newtheorem{theorem}{Theorem}[section]
\newtheorem{algorithm}[theorem]{Algorithm}

\newcommand{\bew}{\noindent\underline{Proof.}\ }

\newtheorem{remark}[theorem]{Remark}

\newtheorem{proposition}[theorem]{Proposition}
\newtheorem{definition}[theorem]{Definition}
\newtheorem{defn}[theorem]{Definition}

\newcommand{\fa}{\mbox{ for}\mbox{ all }}

\newcommand{\Z}{{\mathbb{Z}}}

\newcommand{\F}{{\mathbb{F}}}

\newcommand{\A}{{\mathbb{A}}}

\newcommand{\C}{{\mathbb{C}}}
\newcommand{\trace}{\mbox{trace}}
\newcommand{\tr}{\textup{\scriptsize tr}}
\newcommand{\eb}{\phantom{zzz}\hfill{$\square $}\smallskip}

\title{Degenerate flag varieties in network coding}
\author{Ghislain Fourier and  Gabriele Nebe} 
 
\begin{document}
\maketitle
\textsc{
RWTH Aachen University,
52056 Aachen, Germany}

\emph{E-mail address}{:\;\;}
\texttt{fourier@mathb.rwth-aachen.de}, 
\texttt{nebe@math.rwth-aachen.de}
\medskip

{\sc Abstract.} 
Building upon the application of flags to network coding introduced 
 in \cite{LNV}, we develop a variant of this coding technique that 
 uses degenerate flags. 
 The information set is a metric affine space isometric to 
 the space of upper triangular matrices endowed with the 
 flag rank metric. This suggests the development of a 
 theory for flag rank metric codes in analogy to the rank metric
 codes used in linear subspace coding.



\bigskip

\section{Introduction}
\label{intro}

Linear subspace coding is  a well established 
 theory to transmit information in large unknown networks (see for instance
\cite{SKK}, \cite{KK}).
 Here information is encoded in a subspace $U$ of the rowspace $V:=K^{n+1}$, 
 of rows of length $(n+1)$ over some field $K$.
The source successively inputs vectors from $U$ and in network nodes
forward $K$-linear combinations of the received vectors.
 The success of the invention by Silva, K\"otter, and Kschischang 
 is partly based  on the fact that 
 coding for errors and erasures 
 can be done in a linear space leading to techniques that are
 very similar to the classical block coding techniques. 
 The idea here is to work with subspaces of a fixed dimension, say $i$, 
 and hence in the Grassmannian 
 $$\Gr_i(V) := \{ U\leq V \mid \dim (U )  = i \} , $$
 a full orbit under right multiplication with $\GL_{n+1}(K)$. 
 The orbit $\Gr_i(V)$ is partitioned further into so called {\em cells},
 according to the pivot positions of the subspace.  
 Each cell is a metric  affine space
 with translations preserving the Grassmann distance. 
 In particular the largest cell 
 $$\Gr_i^{(0)} (V) := \{ U \in \Gr _{i}(V) \mid U = U_A = \textup{ row space } (I_i | A) 
 \textup{ for some } A\in K^{i\times (n+1-i)}  \} $$
 is a regular orbit of the unipotent abelian subgroup 
 $$R_i = \{ \left( \begin{array}{cc} I_i & A \\ 0 & I_{n+1-i} \end{array} \right) \mid A\in K^{i\times (n+1-i)} \} \leq \GL _{n+1}(K) .$$  

	 To send the matrix $A\in K^{i\times (n+1-i)}$ through the network, 
 the source inputs the rows of $(I_i | A )$. In network nodes 
 forward linear combinations of the received vectors. 
 So the information preserved by the network is the subspace $U_A \in \Gr_i^{(0)}(V)$. The receiver gets the rows of $(G| GA )$, where $G\in K^{s\times i}$  will very likely have rank $i$ for $s$ big enough. 
 Using Gaussian elimination, the receiver computes the row reduced 
 echelon form $(I_i|A) $ of $(G|GA)$ to recover the matrix $A$. 

 To measure errors and erasures one uses the {\em Grassmann distance} 
 $$d(U,W):= d_i (U,W) := \frac{1}{2} (\dim (U+W) - \dim (U\cap W) ) \mbox{ for all } U,W \in \Gr _i(V) .$$
 This is a $\GL_{n+1}(K)$-invariant distance, so also the matrices in the 
 subgroup $R_i$ above preserve distances. 
 More precisely we have 
 $d(U_A,U_B) = \rk (A-B )$ for all $A,B \in K^{i\times (n+1-i)}$, 
 so the cell above 
 is isometric to the vector space 
 $K^{i\times (n+1-i)}$ endowed with the rank metric.
 This observation motivated a very successful study
 of rank metric codes (see for instance \cite{Ravagn} and the 
 references in this paper).

 In the paper \cite{LNV} we suggest a new feature to network 
 coding which we call linear flag coding. 
 The idea 
 origins from observing the behaviour of a commonly used network, the line network, 
 where all nodes are aligned. In this case the linear combination 
 of the vectors forwarded by an in network node only involves vectors 
 that are sent before, so the network automatically preserves the flag
 defined by the sent vectors. 
Also  in information transmission
 via unknown networks, packages need to be given a sequence number, 
 as they usually do not arrive at the receiver in the same order as they
 are sent. In \cite{LNV} we developed basic techniques for using 
 flags in network coding. The idea is that the source inputs, 
 as in linear network coding, a sequence of vectors  $u_i$.
 Now the in network nodes only perform 
 linear combinations with vectors having a lower sequence
 number.
	 So the information that travels through the network is 
	 the flag  $$ (U_1,\ldots , U_{n}) \in \Fl (V)  $$
where $U_i = \langle u_1,\ldots , u_i \rangle $ for all $i$. 
The flag variety 
$$\Fl(V) := \{ (U_1,\ldots , U_{n}) \in \prod _{i=1}^{n} \Gr_i(V) \mid 
	 U_1\leq U_2\leq \ldots \leq U_{n} \} $$ 
	 is again a full orbit under simultaneous right multiplication 
	 by $\GL_{n+1}(K)$. Natural $\GL_{n+1}(K)$-invariant distances 
	 are studied in \cite{Liebhold} and \cite{LNV}, where we propose
	 the sum of the Grassmann distance as a suitable distance function 
	 ${\bf d}$ on $\Fl(V)$.
Again $\Fl(V)$ partitions into cells. These cells 
are regular orbits under certain unipotent algebraic subgroups
	 of $\GL_{n+1}(K)$; the largest 
	 cell 
	 $$\Fl^{(0)}(V) :=  \Fl(V) \cap  
 \prod _{i=1}^{n} \Gr_i^{(0)}(V) $$
	 has dimension $n(n+1)/2$. One major disadvantage here is that 
	 the relevant subgroups are not abelian. 

	 The present paper aims to overcome this difficulty by replacing 
	 the flag variety by a certain natural degeneration, $\Fl ^{(a)}(V)$
	 (see Definition \ref{Grassdist}). 

	 In mathematical theory, 
the degenerate flag variety $\Fl^{(a)}(V)$
has been introduced in the framework of PBW degenerations by E.~Feigin \cite{Fei11}. 
In fact, the ordinary flag variety is isomorphic to the highest weight orbit of
any regular, finite-dimensional $\operatorname{GL}_{{n+1}}(K)$-module $L_\lambda$.  
The PBW degeneration $L_\lambda^a$ of $L_\lambda$ is induced by the natural PBW 
filtration on the universal enveloping algebra $U(\mathfrak{gl}_{{n+1}}(K))$ 
\cite{FFL10}. $L_\lambda^a$ is again a regular module for an (abelian) 
algebraic group of dimension $n(n+1)/2$. 
Then E.~Feigin defined the PBW degenerated flag variety as the highest weight orbit 
(of an abelian algebraic group) on $L_\lambda^a$. 
Hence, by construction, $\Fl^{(a)}(V)$
is a projective variety of dimension $n(n+1)/2$ over $K$. 
Again in \cite{Fei11}, a realization of the degenerate flag variety
in terms of subspace conditions is provided, the one we will be working with for our purposes. 

For applications it is enough to work with the easily accessible 
linear algebra model 
given in \cite{Lanini}.
We obtain a  cell decomposition of $\Fl^{(a)}(V)$ as for flags 
where the largest cell 
$$\Fl^{(a0)}(V) = \Fl^{(a)}(V) \cap \prod _{i=1}^{n} \Gr_i^{(0)}(V) $$  again has dimension $n(n+1)/2$.
	 The use of degenerate flags  has two advantages:
	 On the one hand the manipulations by the in network nodes 
	 require less additions compared to the linear 
	 subspace or flag coding (see Remark \ref{cost}). 
	Much more important: the cells in  $\Fl ^{(a)}(V)$ are again 
	 affine spaces yielding an even better analogy between degenerate 
	 flag coding and linear
	 subspace coding. 
 In particular the largest cell $\Fl^{(a0)}(V)$ is a regular orbit of the 
	 vector space $U^{n}(K)$ 
	 of all $n\times n $ upper triangular matrices.
	 The space $U^{n}(K)$ has dimension $n(n+1)/2$.
	 As the translations preserve the
	  Grassmann distance,
	  this distance 
translates to a generalised rank metric
(see Definition \ref{flagrk}) 
on $U^{n}(K)$
which we call the {\em flag rank metric}.
Flag rank metric codes are introduced in Section \ref{frkcodes}. 
In particular we elaborate codes in which each two distinct elements 
have maximal possible distance (Propositions \ref{evenmaxdist} and \ref{oddmaxdist}). A more sophisticated analysis of flag rank metric codes 
is subject to further research. 
We give the network coding scheme using degenerate flags in Section \ref{network} before we conclude the paper with a comparison of some network coding
properties of degenerate flags 
and flags. 

{\sc Acknowledgements:} We thank Prof. Dr. Angeles Vazquez-Castro for many
comments that helped to improve this paper.
The authors gratefully acknowledge support by DFG-collaborative research center TRR 195.

\section{Degenerate flag varieties} 

Throughout the paper we 
let $K$ be a field and $V$ be a vector space over $K$ of dimension $n+1$. 
The {\bf Grassmann} variety 
$$\Gr_i(V) := \{ U \leq V \mid \dim (U ) = i \} $$ 
is the set of $i$-dimensional subspaces of $V$. It becomes a metric space by putting 
 $$d(U,W):= d_i (U,W) := \frac{1}{2} (\dim (U+W) - \dim (U\cap W) ) \mbox{ for all } U,W \in \Gr _i(V) ,$$
the Grassmann distance on $\Gr_i(V)$. 

\begin{remark}\label{maxdistGr}
	The maximum distance on $\Gr_i(V)$ is
	$$\max \{ d_i (U,W) \mid U,W\in \Gr_i (V) \} = \min \{ i, n+1-i \} .$$
\end{remark}

\bew
If $2i \leq n+1$ then we can find $U,W\in \Gr_i(V) $ such that $U\cap W = \{ 0 \}$. 
Then $\dim (U+W) = \dim(U) + \dim (W) = 2i = 2 d_i(U,W) .$
If $2i > n+1$ then we can find $U,W\in \Gr_i(V) $ such that $U+W = V$. 
Then $\dim (U\cap W) = 2i - (n+1) $ and hence $d_i(U,W) = n+1-i$.
\eb

Full flags and also the degenerate flags that we consider in the present paper 
are subvarieties of the direct product of Grassmannians. 
The sum of the Grassmann distances defines a metric on this direct product:

\begin{remark}\label{maxdistpGr}
	The function
	 $${\bf d}: \prod _{i=1}^n \Gr_i(V) \times \prod _{i=1}^n \Gr_i(V) \to \Z _{\geq 0} , \
	{\bf d} ((U_1,\ldots , U_n), (V_1,\ldots , V_n)) := 
	\sum _{i=1}^n d_i(U_i,V_i) $$ 
defines a metric on the product of the Grassmannians which we call 
	again {\em Grassmann metric}.
	The maximum distance between two elements in $\prod _{i=1}^n \Gr_i(V)$
	is 
	$$\max \{ {\bf d} ({\bf U},{\bf W}) \mid {\bf U},{\bf W}\in \prod _{i=1}^n \Gr_i(V) \} = {\bf d}_{max} (n+1)  $$
	with 
	$${\bf d}_{max} (n+1) := \left\{ \begin{array}{ll} k^2 & n+1 = 2k \textup{ even} \\ 
k(k+1) & n+1 = 2k+1 \textup{ odd.} \end{array} \right.  $$
\end{remark}

\bew
It is immediate that the sum ${\bf d}$ is again a metric. 
So we only show the upper bound. 
First assume that $n+1 = 2k$ is even. 
Then for any $U_i,V_i \in \Gr_i(V)$ ($i=1,\ldots , n$) we have 
$d_i(U_i,V_i) \leq \min \{ i, 2k -i \} $ by Remark \ref{maxdistGr}.
So 
$$\sum _{i=1}^{2k-1} d_i(U_i,V_i) \leq \sum _{i=1}^k i + \sum _{i=k+1}^{2k-1} (2k-i) = 2 \sum _{i=1}^{k-1} i + k = k(k-1) +k = k^2.$$ 
Now assume that $n+1 = 2k+1$ is odd. 
Then for any $U_i,V_i \in \Gr_i(V)$ ($i=1,\ldots , n$) we have 
$d_i(U_i,V_i) \leq \min \{ i, 2k+1 -i \} $ (again by Remark \ref{maxdistGr}). 
So 
$$\sum _{i=1}^{2k} d_i(U_i,V_i) \leq \sum _{i=1}^k i + \sum _{i=k+1}^{2k} (2k+1-i) = 2 \sum _{i=1}^{k} i  = k(k+1) .$$ 
\eb

To define the degenerate flag variety we use the notation of \cite{Lanini}. 
Fix a basis $(f_1,\ldots ,f_{n+1})$ of $V$ and identify the elements of $V$ 
with their coordinate rows,  $V\cong K^{1\times (n+1)}$. 
For $1\leq i \leq n+1$ we define the projection 
$$\pr _i : V \to V , f_j \mapsto \left\{ \begin{array}{ll} f_j & \mbox{ if } j\neq i \\
0 & \mbox{ if } j=i \end{array} \right. .$$
With respect to the chosen basis, the endomorphism $\pr _i$ acts by 
right multiplication with the diagonal matrix 
$$D_i = \diag(1,\ldots , 1, 0 ,1,\ldots , 1) \in K^{(n+1)\times (n+1)}$$ having the $0$ on position $(i,i)$. 
For $j < i$ we put $\pr_{j,i} := \prod _{k=j+1}^{i} \pr _{k} $ and 
$$D_{j,i} = \prod _{k=j+1} ^i D_k = \diag(1,\ldots, 1, 0,\ldots, 0, 1\,\ldots , 1),$$ with in total $i-j$ zeros on the
diagonal on positions $(j+1,j+1), \ldots , (i,i) $.
 In abuse of notation we also put $\pr _{i,i} := \id _V$.
 Then clearly
$$\rk (\pr_i) = \rk (D_i) = n ,\ \pr _j \pr_i = \pr_i \pr _j ,\ \rk (\pr _{j,i} ) =\rk (D_{j,i}) = n-i+j .$$

 \begin{defn}\label{Grassdist}
	 The degenerate flag variety $\Fl ^{(a)}(V)$ is defined as
	 $$\Fl ^{(a)} (V) := \{ (V_1,\ldots , V_n) \in \prod _{i=1}^n \Gr_i(V)  \mid \pr _{i+1}(V_i) \subseteq V_{i+1} \fa i=1,\ldots , n-1\} .$$
	 $\Fl ^{(a)}(V) $ becomes a metric space via
	 the restriction of the Grassmann distance ${\bf d}$ 
	 from Remark \ref{maxdistpGr}.
\end{defn} 

From Remark \ref{maxdistpGr} we immediately conclude 

\begin{remark}\label{maxdistflag}
	The maximum distance on $\Fl^{(a)}(V)$ is bounded from
	above by ${\bf d}_{max} (n+1) $.
\end{remark}

As we will see below (Proposition \ref{evenmaxdist} and Proposition \ref{oddmaxdist}) the maximum distance ${\bf d}_{max}(n+1) $ is achieved in 
 the subset $\Fl^{(a0)}(V)$ of $\Fl^{(a)}(V)$. 

The main result  of \cite{Lanini} establishes a torus invariant 
bijection between a slight variation of 
 the metric space $\Fl ^{(a)}(V)$ and a partial flag variety.
In particular \cite[Theorem 1.2]{Lanini} elaborates an isometry between 
a certain Schubert cell in this partial flag variety and 
a cell $\Fl^{(a0)}(V)$ of maximal dimension in $\Fl ^{(a)}(V)$. 
It turns out that this cell is indeed an affine space of 
dimension $n(n+1)/2$ where the translations respect the Grassmann 
metric ${\bf d}$ (see Theorem \ref{frk}). 

To define this cell $\Fl^{(a0)}(V)$ we fix the 
 basis $(f_1,\ldots , f_{n+1} ) $ of $V$ from above 
 with which we identify 
 $V$ with the row space $K^{1\times (n+1)} $. 
 Recall that $$\Gr _i^{(0)}(V)  = \{ U_A = 
 \textup{ row space } (I_i | A) \mid 
 A\in K^{i\times (n+1-i)}  \} .$$

 \begin{defn}\label{Fla0}
	 $\Fl^{(a0)} (V):=  \Fl^{(a)}(V) \cap \prod _{i=1}^n \Gr _i^{(0)} (V) $.
 \end{defn}

Given $(V_1,\ldots , V_n) \in \Fl^{(a0)} (V)$ then 
for any $i$ the space $V_i=U_{A_i}$ is the row space of
$$\Psi_i (A_i)   := 
\left( \begin{array}{c|c}  I_{i}  &
	A_i  \end{array}  \right) \in K^{i\times (n+1)} $$
for a unique matrix $A_i = A(V_i)  \in K^{i\times (n+1-i) }$. 
The condition that $\pr _{i+1} (V_i) \subseteq V_{i+1}$ 
is equivalent to the condition that 
$$A(V_{i+1}) = \left( \begin{array}{c} \widehat{A(V_i)} \\ a_i \end{array} \right)  \textup{ for some row } a_i \in K^{1 \times (n-i) } $$
where $\widehat{A_i} $ is the $i\times (n-i)$ matrix 
obtained by deleting the first column from the $i\times (n+1-i) $ matrix $A_i$.

For an $n\times n$-matrix $\Delta \in K^{n\times n}$ and $1\leq i\leq n$ 
we denote by 
$$\Delta _{[i]} := (\Delta _{k,\ell } )_{k=1,\ldots , i, \ell = i,\ldots n } \in K^{i\times (n+1-i)} $$ the upper right rectangular corner of $\Delta $. 
Using this notation we obtain the following parametrization of 
$\Fl^{(a0)}(V)$ by upper triangular matrices:

\begin{defn}\label{Delta}
The space of upper triangular matrices is denoted by 
$$U^n(K) := \{ \Delta \in K^{n\times n} \mid \Delta _{ij} = 0 \textup{ for } i>j \} .$$
	The function
	$\Delta ^{(a)} : \Fl ^{(a0)} (V)\to U^n(K) $ 
	associates to   $(V_1,\ldots , V_n) \in \Fl^{(a0)} (V) $ 
	the upper triangular matrix 
	$$\Delta := \Delta ^{(a)}  ((V_1,\ldots , V_n)) \in K^{n\times n }  $$ 
	defined by $\Delta _{[i]} := A(V_i) $. 
	It defines a bijection between $\Fl ^{(a0)}(V)$ and $U^n(K)$ where 
the inverse mapping is given by 
	$$ {\mathfrak F}^{(a)} : U^{n}(K) \to \Fl ^{(a0)}(V) , 
\Delta \mapsto (V_1,\ldots  , V_n)  $$ 
where $V_i = U_{\Delta _{[i]}} $ is the row space of $ \Psi_i (\Delta _{[i]} ) $ for all $i$.
\end{defn}

To obtain an isometry with respect to the Grassmann distance ${\bf d} $ from
Definition \ref{Grassdist} we define the flag rank metric $\frk $ on $U^n(K)$. 

\begin{defn}\label{flagrk}
	The {\em flag rank} of a matrix $\Delta \in U^n(K) $ is defined as 
	$$\frk (\Delta ) := \sum _{i=1}^n \rk (\Delta _{[i]} ) .$$
\end{defn}

\begin{theorem} \label{frk}
	For $(U_1,\ldots, U_n), (V_1,\ldots , V_n) \in \Fl ^{(a0)}(V) $ 
	we have
	$${\bf d} ((U_1,\ldots, U_n), (V_1,\ldots , V_n) ) = 
	\frk ( \Delta^{(a)} (U_1,\ldots, U_n) - \Delta^{(a)} (V_1,\ldots , V_n) ) .$$
\end{theorem}

\bew
Let $\Delta := \Delta^{(a)} (U_1,\ldots ,U_n) $ and 
 $\Lambda  := \Delta^{(a)} (V_1,\ldots ,V_n) $.
 Then 
 $U_i$ is the row space of $\psi _i (\Delta _{[i]} ) $ 
 and $V_i$ is the row space of $\psi _i (\Lambda _{[i]} )$ 
 for $1\leq i \leq n$. So by \cite[Proposition 4]{SKK} we have 
 $d_i (U_i,V_i) = \rk (\Delta _{[i]} - \Lambda _{[i]} )$.
 Summing over all $i$ yields the result.
\eb

We conclude this section with an algorithm for computing $\Delta ^{(a)}((U_1,\ldots , U_n) )$ for any $(U_1,\ldots , U_n) \in {\Fl }^{(a0)} (V) $:

\begin{algorithm} \label{algDelta} 
	{\bf Input:} $(U_1,\ldots , U_n) \in {\Fl }^{(a0)} (V) $. \\
	{\bf Output:} $\Delta  := \Delta^{(a)} ((U_1,\ldots , U_n) ) \in U^n(K)$. \\
	{\bf Algorithm:} 
	\begin{itemize}
	\item[(1)] Choose $u = (u_1,\ldots , u_{n+1}) \in U_1$ such that 
		$u_1\neq 0 $ and put $\Delta _{[1]} := (u_2/u_1,\ldots , u_{n+1}/u_1 ) $. 
	\item[(2)] Assume that we computed $\Delta _{[i-1]} \in K^{(i-1) \times  (n-i+2) }$ and put 
		$A:=   \widehat{\Delta _{[i-1]}}  \in K^{(i-1) \times (n-i+1)} $
			by deleting the first 
			 column of  
			 $\Delta _{[i-1]}$.
	\item[(3)]
		Choose $w = (w_1,\ldots , w_{n+1}) \in U_i $ such that 
			$w_i \neq 0 $. 
			Put $v := (w_{i+1} , \ldots , w_{n+1}) \in K^{n-i+1} $
			and compute 
			$$u:= v - \sum _{j=1}^{i-1} w_j A_j $$ 
			where $A_j$ is the $j$-th row of the matrix $A$ from 
			(2). 
			Put 
			$$\Delta _{[i]} := \left( \begin{array}{c} 
				A \\ 
				(u_{i+1}/w_i, \ldots , u_{n+1}/w_i) 
			\end{array} \right) .$$
		\item[(4)] Return the matrix $\Delta \in U^{(n)}(K)$ 
			obtained from the computed $\Delta _{[i]}$, 
			$i=1,\ldots , n$.
	\end{itemize}
\end{algorithm} 

\section{Flag rank metric codes in $U^n(K)$} \label{frkcodes}

\begin{defn} 
	A {\em flag rank metric code} ${\mathcal C}$ is a $K$-subspace 
	${\mathcal C} \leq U^n(K)$. 
	The {\em minimum distance} of ${\mathcal C}$ is defined as 
	$$d_{\frk }({\mathcal C} ) := \min \{ \frk (\Delta ) \mid 
	0\neq \Delta \in {\mathcal C} \} .$$
\end{defn} 

Remark \ref{maxdistflag} and Theorem \ref{frk} hence imply

\begin{remark}\label{maxfrk} 
	$\max \{ \frk (\Delta ) \mid \Delta \in U^n(K) \} \leq {\bf d}_{max}(n+1)$.
\end{remark}

\begin{proposition} \label{evenmaxdist}
	Assume that $n+1 = 2k$ is even. 
	\begin{itemize}
		\item[(a)]  $ \max \{ \frk (\Delta ) \mid \Delta \in U^n(K) \} = k^2 = {\bf d}_{max}(n+1) $.
		\item[(b)] For any flag rank metric code 
			${\mathcal C} \leq U^n(K) $ with 
			$d_{\frk }({\mathcal C} ) = k^2$ we have 
			$\dim ({\mathcal C} ) \leq k$. 
		\item[(c)] If $K$ admits a field extension of degree $k$ then
			there is a flag rank metric code 
			${\mathcal C} \leq U^n(K) $ with 
			$$d_{\frk }({\mathcal C} ) = k^2 \textup{ and } 
			\dim ({\mathcal C} ) = k .$$
	\end{itemize} 
\end{proposition}

\bew
To see (c) let $F$ be a field extension of $K$ of degree $k$. 
The regular representation 
$\rho : F \hookrightarrow K^{k\times k} $ defines an embedding of 
$F$ into $K^{k\times k}$ 
such that for any $0\neq f\in F $ the 
matrix $\rho(f)$ is invertible, so has full rank $k$. 
Now let 
$${\mathcal C} := \left\{ \Delta _f:=\left( \begin{array}{cc} 0 & \rho(f) \\ 0 & 0 \end{array} 
\right) \mid f \in F \right\} \leq U^n(K).$$
Note that the $0$s on the diagonal are $(k-1) \times k$ respectively $k\times (k-1)$-matrices. 
Then clearly $\dim ({\mathcal C}) = \dim (F) = k$.
As any $\ell $ rows or columns of $\rho(f)$ are linearly independent for any $0\neq f\in F$ we compute 
$d_{\frk} (\Delta _f) = {\bf d}_{max}(2k) $. \\
Now (a) 
 follows from the fact that  $\frk(\Delta _1) = {\bf d}_{max}(n+1)$ together with
 Theorem \ref{frk} and Remark \ref{maxdistflag}. 
\\
To prove (b) let ${\mathcal C} \leq U^{n}(K) $ be a flag rank metric code
of dimension $\geq k+1$. 
Then the projection of ${\mathcal C}$ onto the first $k$ entries of the
last column is not injective, so ${\mathcal C}$ contains some matrix 
$M$ of the form 
$$ M = \left( \begin{array}{cc} X  & 0 \\ 
0 & a \end{array} \right) \textup{ with } X\in K^{k\times (2k-2) }  .$$
If $Y\in K^{k\times (k-1)} $ consists of the last $k-1$ columns of 
$X$ then $M_{[k]} = \left(\begin{array}{c|c} Y & 0 \end{array} \right) \in K^{k\times k} $ has at most rank $k-1$, so $\frk(M) \leq k^2-1 $. 
\eb

\begin{proposition} \label{oddmaxdist}
	Assume that $n+1 = 2k+1$ is odd. 
	\begin{itemize}
		\item[(a)]  $ \max \{ \frk (\Delta ) \mid \Delta \in U^n(K) \} = k(k+1) = {\bf d}_{max}(n+1) $.
		\item[(b)] For any flag rank metric code 
			${\mathcal C} \leq U^n(K) $ with 
			$d_{\frk }({\mathcal C} ) = k(k+1)$ we have 
			$\dim ({\mathcal C} ) \leq k+1$. 
		\item[(c)] If $K$ admits a field extension of degree $k$ then
			there is a flag rank metric code 
			${\mathcal C} \leq U^n(K) $ with 
			$$d_{\frk }({\mathcal C} ) = k(k+1) \textup{ and } 
			\dim ({\mathcal C} ) = k .$$
	\end{itemize} 
\end{proposition}

\bew
To see (c) we consider the same code 
$${\mathcal C} := \left\{ \Delta _f:=\left( \begin{array}{cc} 0 & \rho(f) \\ 0 & 0 \end{array} 
\right) \mid f \in F  \right\} \leq U^n(K)$$
as in the proof of Proposition \ref{evenmaxdist}.
Now the $0$s on the diagonal are $k \times k$-matrices. 
As before the minimum distance of ${\mathcal C}$ is $k(k+1)$ and 
 $\dim ({\mathcal C}) = k$.
 \\
Again (a) 
 follows from the fact that  $\frk(\Delta _1) = {\bf d}_{max}(n+1)$ together with
 Theorem \ref{frk} and Remark \ref{maxdistflag}. 
\\
To prove (b) let ${\mathcal C} \leq U^{n}(K) $ be a flag rank metric code
of dimension $\geq k+2$. 
Then the projection of ${\mathcal C}$ onto the last $k+1$ entries of the
first row is not injective, so ${\mathcal C}$ contains some matrix 
$M$ of the form 
$$ M = \left( \begin{array}{cc} a  & 0 \\ 
b & X  \\ 0 & y \end{array} \right) \textup{ with } X\in K^{(k-1)\times (k+1) }  .$$
Then $M_{[k]} = \left(\begin{array}{c} 0  \\ X \end{array} \right) \in K^{k\times (k+1)} $ has at most rank $k-1$, so $\frk(M) \leq k(k+1)-1 $. 
\eb

\subsection{An example} 
To illustrate the notation we give a small example.
	Let $K=\F_3 $ be the field with 3 elements and let $n=4$. 
	Let ${\mathcal T}$ be the 4-dimensional flag rank metric code with basis 
	$$\left\{ 
	\left( \begin{array}{cccc} 
1&0&1&1\\
0&1&0&0\\
0&0&1&1\\
0&0&0&0 \end{array} \right) , 
\left( \begin{array}{cccc} 
0&2&1&0\\
0&2&2&0\\
0&0&1&0\\
0&0&0&1 \end{array} \right) , 
\left( \begin{array}{cccc} 
0&0&1&0\\
0&0&0&1\\
0&0&0&0\\
0&0&0&0 \end{array} \right) , 
\left( \begin{array}{cccc} 
0&0&0&1\\
0&0&2&0\\
0&0&0&0\\
0&0&0&0 \end{array} \right)  \right\}. $$ 
	Then 
 ${\mathcal T}\leq U_4(K)$ is a flag rank metric code of dimension 4 
	with minimum distance  5. 
	This is best possible for any field $K$: 
	If ${\mathcal C}  \leq U_4(K) $ is 
	a flag rank metric code with minimum distance $\geq 5$, then 
	the intersection of ${\mathcal C}$  with the 6-dimensional 
	subspace 
	$${\mathcal D}:= \left( \begin{array}{cccc} K & K & 0 & 0 \\ 0 & K & 0 & 0 \\
	0 & 0 & K & K \\ 0 & 0 & 0 & K \end{array} \right) $$ 
	is $\{ 0 \}$ as the maximal flag rank of a matrix in ${\mathcal D}$ is 
	4. So the dimension of ${\mathcal C}$ cannot exceed 
	$\dim (U_4(K)) - \dim ({\mathcal D}) = 10 -6 = 4$.

	In the next table we apply the degenerate flag coding map 
	${\mathfrak F}^{(a)} $ to the first two basis elements and their sum:

	$$
	\begin{array}{c|l} 
		\Delta & {\mathfrak F}^{(a)}(\Delta )=(U_1,U_2,U_3,U_4) \\
		\hline
		& \\
 \left( \begin{array}{cccc} 1&0&1&1\\
0&1&0&0\\
0&0&1&1\\
0&0&0&0 \end{array} \right) 
		& \begin{array}{l} 
U_1 = \langle f_1+f_2+f_4+f_5 \rangle   \\
U_2 = \langle f_1 + f_4 + f_5, f_2 + f_3 \rangle   \\
U_3 = \langle f_1+f_4+f_5, f_2,  f_3+f_4+f_5 \rangle   \\
			U_4 = \langle f_1+f_5, f_2,  f_3+f_5, f_4 \rangle  
		\end{array} \\

		& \\
\left( \begin{array}{cccc} 
0&2&1&0\\
0&2&2&0\\
0&0&1&0\\
0&0&0&1 \end{array} \right)  & 
		\begin{array}{l} 
U_1 = \langle f_1+2f_3+f_4 \rangle   \\
U_2 = \langle f_1 + 2f_3 + f_4 , f_2 + 2f_3 +2f_4 \rangle   \\
U_3 = \langle f_1+f_4, f_2+2f_4,  f_3+f_4 \rangle   \\
		U_4 = \langle f_1, f_2,  f_3, f_4+f_5 \rangle \end{array} \\
		& \\
\left( \begin{array}{cccc} 
1&2&2&1\\
0&0&2&0\\
0&0&2&1\\
0&0&0&1 \end{array} \right) 
		& 
		\begin{array}{l}
U_1 = \langle f_1+f_2+2f_3+2f_4+f_5 \rangle   \\
U_2 = \langle f_1 + 2f_3 + 2f_4 +f_5, f_2  +2f_4 \rangle   \\
U_3 = \langle f_1+2f_4+f_5, f_2+2f_4,  f_3+2f_4 +f_5\rangle   \\
	U_4 = \langle f_1+f_5, f_2,  f_3+f_5, f_4+f_5 \rangle  \end{array} 
	\end{array}
	$$

\subsection{Duality and decoding of flag rank metric codes} 

The space $U^n(K)$ is equipped with a non degenerate $K$-bilinear form
$$ A\cdot B := \trace (A B^{\tr }) \textup{ for all } A,B\in U^n(K).$$

We hence have the notion of dual code ${\mathcal C}^{\perp }$ of
a flag rank metric code and can use syndrome decoding to find an element of the
flag rank metric code ${\mathcal C} \leq U^n(K)$ having minimum distance to 
any given element $\Delta \in U^n(K) $. 

\begin{definition}
	For ${\mathcal C}\leq U^n(K) $
	 the {\em dual code} is 
	$${\mathcal C}^{\perp } := \{ \Delta \in U^n(K) \mid \trace (\Delta  \Lambda ^{\tr }) = 0 \mbox{ for all } \Lambda \in {\mathcal C} \} .$$
\end{definition}

From the non-degeneracy of the bilinear form we obtain that 
$$\dim ({\mathcal C} ) + \dim ({\mathcal C}^{\perp} ) = \dim (U^n(K)) = n(n+1)/2 \mbox{ and } 
{\mathcal C} = ({\mathcal C}^{\perp } ) ^{\perp } .$$

We now fix a flag rank metric code ${\mathcal C} \leq U^n(K) $ and 
choose a basis  $\underline{\Delta }:= (\Delta _1,\ldots , \Delta _{\ell } )  $  of ${\mathcal C}^{\perp }$. 

\begin{remark} \label{properties}
	\begin{itemize}
		\item[(a)]
	For $A\in U^n(K)$ the {\em syndrome} of $A$ with respect to $\underline{\Delta }$ is 
	$$\syn_{\underline{\Delta}} (A) := ( \trace (\Delta _1 A^{\tr} ) ,\ldots , \trace (\Delta _{\ell } A^{\tr} ) ) \in K^{\ell } .$$
\item[(b)] For $A\in U^n(K) $ we have that $A\in {\mathcal C}$ if and only if $\syn_{\underline{\Delta}} (A) = (0,\ldots , 0 )$.
\item[(c)]
	For  $f\in K^{\ell }$ the set 
	$$E_f := \{ A\in U^n(K) \mid \syn_{\underline{\Delta}} (A) = f \} = A_f + {\mathcal C}$$ 
	is a coset of ${\mathcal C}$ in $U^n(K)$. 
\item[(d)]
	We have that $E_{af} = a E_f = aA_f + {\mathcal C} $ for any $a\in K$.
\item[(e)] 
	Let $d_f := \min \{\frk (A) \mid A \in E_f \} $. Then any $A_f\in E_f$ with $\frk (A_f) = d_f$ is called 
	a minimal coset leader for $f$. 
\item[(f)] 
	For $0\neq a\in K$ and $f\in K^{\ell }$ we have that $d_{af} = d_f$.
	If $A_f$ is a minimal coset leader for $f$ then $aA_f$ is a minimal coset leader for $af$.
\end{itemize}
\end{remark}
	
The usual syndrome decoding algorithm for the case that $K$ is a finite field now reads as follows:

\begin{algorithm}\label{decfrk}
	{\bf Input:} A flag rank metric code ${\mathcal C} \leq U^n(K)$ and $A\in U^n(K)$. \\
	{\bf Output:} One element $C\in {\mathcal C}$ such that $\frk (A-C) = \min \{\frk (A-C') \mid C'\in {\mathcal C} \} $.
	\\
	{\bf Precomputing:} 
	\begin{itemize} 
		\item[P1] 
			Compute a basis  $\underline{\Delta }:= (\Delta _1,\ldots , \Delta _{\ell } )  $  of ${\mathcal C}^{\perp }$.
		\item[P2] For all $\langle f \rangle \in \Gr _1 (K^{\ell }) $ compute a minimal coset leader $A_f$ for $f$ in $E_f$ 
			and store the set $M_{\underline{\Delta}} := \{ (f,A_f) \mid \langle f\rangle \in \Gr_1(K^{\ell}) \}$.
	\end{itemize}
	{\bf Decoding:} 
	\begin{itemize}
		\item[D1]
			Compute $\syn_{\underline{\Delta}} (A) =: g$. If $g= (0,\ldots ,0)$ then return $A$. 
\item[D2] Find the $(f,A_f) \in M_{\underline{\Delta }} $ such that $\langle f \rangle = \langle g \rangle $ and $a\in K$ with $g=af$.
\item[D3] Return $A-aA_f$. 
	\end{itemize}
\end{algorithm}

\bew
We have $C:=A-aA_f \in {\mathcal C}$ by Remark \ref{properties} (b) 
as $\syn_{\underline{\Delta }}(A-aA_f) = g-af = 0 $.
Moreover 
$$\frk (A-C) = \frk (aA_f) = \frk (A_f) =  d_f = d_g =
\min \{ \frk (A-C') \mid C'\in {\mathcal C} \}  .$$
\eb

The syndrome decoding algorithm is a very general,  in general not very efficient, decoding algorithm. 
For special types of flag rank metric codes, there are much better decoding procedures as illustrated for 
flag codes in \cite[Section 3.4.4]{Liebhold}. 

\section{Network coding with degenerate flags}\label{network}

The network coding techniques from  \cite{LNV} can be easily 
adapted to degenerate flags: 
Any element   ${\bf U} = (U_1,\ldots , U_n) \in {\Fl} ^{(a0)}(V)  $ is represented by a matrix $X  \in K^{n\times (n+1)} $ 
as follows. If $X_i$ is the $i$-th row of $X$, then 
$$U_{i} = \langle  \pr_{j,i} (X_{j}) \mid j=1,\ldots , i \rangle $$
for all $0\leq i < n $.
The source sends the matrix $X$, row by row, where the $i$-th row has sequence number $i$:
$$(X_1,1), (X_2,2) , \ldots , (X_n,n) .$$
In network nodes receive vectors $(Y_i,i)$ for $i=1,\ldots , n$. 
For each $i=1,\ldots ,n$ they form linear combinations 
$$Z_{i} := \sum _{j=1}^{i} a_{ij} \pr_{j,i}(Y_j) \ \ \ \ \ \ (\star )$$ 
where the $(a_{ij})_{1\leq j \leq i \leq n} $ is a lower triangular matrix in 
$K^{n\times n}$. Then the node  forwards $(Z_1,1)$, $\ldots $, $(Z_n,n) $. 
The receiver hence gets a number of rows $(R_{i,t},i) $ for $i=1,\ldots , n$ and $t=1,\ldots ,s_i$ and 
puts 
$$W_{i} := \langle \pr_{j,i}(R_{j,t_j})  \mid
1\leq t_j \leq s_j , 1\leq j \leq i \rangle $$ 
for $i=0,\ldots , n-1$. 
Then ${\bf W} = (W_1,\ldots , W_{n})$ satisfies $\pr _{i+1}(W_i) \subseteq W_{i+1} \fa i=1,\ldots , n-1$.
If $\dim (W_i ) = i$ then we again have that 
${\bf W}  \in {\Fl} ^{{(a)}} (V)$.
As ${\Fl }^{(a0)}(V)$ is dense in ${\Fl}^{(a)}(V)$ it is very likely 
that then ${\bf W} \in {\Fl} ^{(a0)}(V) $. 

\begin{remark}\label{cost}
Note that the operation $(\star )$ for the degenerated flags requires less additions than in 
the network coding scheme that uses full flags. 
If $Y = (y_1,\ldots , y_{n+1}) $, then $\pr_{j,i} (Y) = (y_1,\ldots , y_{j-1}, 0,\ldots , 0, y_{i+1},\ldots , y_{n+1} ) $
so we may just ignore $i-j+1$ entries in $\pr_{j,i} (Y_j) $ when computing $(\star )$.
Assuming that the node received exactly one vector $Y_j$ for each sequence number  $j$, 
the number of coordinate multiplications and additions in time $i+1$ is 
$$\sum _{j=1}^i (n+1-(i-j+1)) = i(n-i) + \frac{i(i+1)}{2} = i(n+1) - \frac{i^2+i}{2} $$ 
compared to $i(n+1)$ in the case of subspace coding or flag coding.
\end{remark}

Using a flag rank metric code ${\mathcal C}\leq U^n(K)$ as input and the precomputation in Algorithm \ref{decfrk} 
we hence get the following algorithm

\begin{algorithm}\label{decflag}
	\begin{itemize}
		\item To send an element $X\in {\mathcal C} $:
		\item the source inputs the $i$-th row $X_i$ of $\psi_i(X_{[i]})$ 
			with sequence number $i$ 
		\item in network nodes receive vectors $(Y_i,i)$ for $i=1,\ldots ,n $ 
		\item compute $Z_{i} := \sum _{j=1}^{i} a_{ij} \pr_{j,i}(Y_j) $ 
			with $a_{ij}\in K$ randomly chosen 
		\item and forward $(Z_{i},i) $ for $=1,\ldots ,n$. 
		\item if $(R_{i,t},i) $,  $t=1,\ldots ,s_i$  are the received vectors 
		with sequence number  $i=1,\ldots , n$ then
	\item the receiver puts
$$W_{i} := \langle \pr_{j,i}(R_{j,t_j})  \mid
1\leq t_j \leq s_j , 1\leq j \leq i \rangle $$ 
for $i=0,\ldots , n-1$. 
\item Check that ${\bf W} = (W_1,\ldots , W_{n}) \in {\Fl} ^{(a0)} (V)$ and compute $A:=\Delta ({\bf W})$ using Algorithm \ref{algDelta}.
\item Use the decoding part of Algorithm \ref{decfrk} to find $C\in {\mathcal C}$ with $\frk (A-C)$ minimal.
\item Decode to $C$.
	\end{itemize}
\end{algorithm}

\section{A comparison} 

This final section highlights the advantages to use degenerate flags 
instead of flags for network coding. 
As before we place ourselves in the situation where $V$ 
is a vector space over a field $K$ of dimension  $(n+1)$. 
Then 
	$$\Fl (V) := \{ (V_1,\ldots , V_n) \in \prod _{i=1}^n \Gr_i(V)  \mid V_i \subseteq V_{i+1} \fa i=1,\ldots , n-1\} $$ and 
	$$\Fl ^{(a)}(V) := \{ (V_1,\ldots , V_n) \in \prod _{i=1}^n \Gr_i(V)  \mid \pr _{i+1}(V_i) \subseteq V_{i+1} \fa i=1,\ldots , n-1\} .$$
Both subsets of 
$\prod _{i=1}^n \Gr_i(V) $ are equipped with the same metric ${\bf d}$, 
the sum of the Grassmann distances. 
Both sets partition into cells according to the pivot sets, where a cell of maximal 
dimension is obtained as 
$$\Fl^{(0)} (V) :=  \Fl(V) \cap \prod_{i=1}^n \Gr_i^{(0)}(V)  \textup{ resp. }
\Fl^{(a0)} (V) :=  \Fl^{(a)}(V) \cap \prod_{i=1}^n \Gr_i^{(0)}(V)  .$$
Both cells have the same dimension and the same maximum distance 
${\bf d}_{max} (n+1)$. 

Both cells are in bijection to 
$U^n(K)$, where for a matrix $\Delta \in U^n(K) $ the degenerate flag 
is ${\mathfrak F}^{(a)}(\Delta ) \in \Fl^{(a0)}(V)$ 
defined after Definition \ref{Delta}. 
For the full flags the map 
$${\mathfrak F}: U^n(K) \to \Fl^{(0)} (V) , 
\Delta \mapsto (U_1,\ldots , U_n) $$ 
where $U_i$ is the space spanned by the first $i$ rows of the 
matrix $\Phi(\Delta ) \in U^{n+1}(K)$ with ones on the diagonal 
and upper triangular entries $\Phi (\Delta )_{i,j+1} = \Delta _{i,j}$ 
for all $1\leq i \leq n, i\leq j \leq n $. 

\begin{remark} 
	$${\bf d} ({\mathfrak F}(0), {\mathfrak F} (\Delta ) ) = \frk (\Delta ) $$ but in general 
	$${\bf d} ({\mathfrak F} (\Gamma ) , {\mathfrak F} (\Delta ) ) = 
	\frk (\Phi^{-1} (\Phi(\Gamma) \Phi (\Delta )^{-1} ) ) $$ 
	involves the non-commutative matrix product of $\Phi (\Delta )$ and $\Phi(\Gamma )$, 
	whereas 
	$${\bf d} ({\mathfrak F}^{(a)} (\Gamma ) , {\mathfrak F} ^{(a)}(\Delta ) ) = 
	\frk (\Gamma - \Delta )  .$$ 
\end{remark}

Also the network coding operations in ${\Fl }^{(a0)}(V) $ 
involve less computations than the ones in ${\Fl } ^{(0)} (V)$ 
(see Remark \ref{cost}). 

In both cases the input alphabet is $U^n(K)$, a vector space of 
dimension $n(n+1)/2$. 
Whereas retrieving the upper triangular matrix $\Delta \in U^n(K)$ 
associated to the flag $ (U_1,\ldots ,U_n) \in {\Fl }^{(0)}(V) $
requires a (partial) Gaussian elimination, 
the decoding algorithm \ref{algDelta} computing 
$\Delta ^{(a)} ((U_1,\ldots , U_n) )$  is much simpler.

\end{document}